\documentclass[]{elsart}

\usepackage{graphicx}
\usepackage{mathrsfs}
\usepackage{amssymb}
\usepackage{bm}

\journal{Physics Letter B}

\begin{document}

\begin{frontmatter}

\title{Single spin asymmetry in $\pi p$ Drell-Yan process}

\author[seu,utfsm]{Zhun Lu},
\author[pku]{Bo-Qiang Ma\corauthref{cor}}
\corauth[cor]{Corresponding author at: School of Physics, Peking
University, Beijing 100871, China.} \ead{mabq@pku.edu.cn},
\author[pku]{Jun She}
\address[seu]{Department of Physics, Southeast University, Nanjing
211189, China}
\address[utfsm]{Departamento de F\'\i sica, Universidad T\'ecnica
Federico Santa Mar\'\i a, and Centro Cient\'\i fico-Tecnol\'ogico de
Valpara\'\i so Casilla 110-V, Valpara\'\i so, Chile}
\address[pku]{School of Physics and State Key Laboratory of Nuclear
Physics and Technology, Peking University, Beijing 100871, China}

\begin{abstract}
We study the single spin asymmetries for the $\pi
p^\uparrow\rightarrow\mu^+\mu^-X$ process. We consider the
asymmetries contributed by the coupling of the Boer-Mulders function
with the transversity distribution and the pretzelosity
distribution, characterized by the $\sin(\phi+\phi_S)$ and
$\sin(3\phi-\phi_S)$ azimuthal angular dependence, respectively. We
estimate the magnitude of these asymmetries at COMPASS by using
proper weighting functions. We find that the $\sin(\phi+\phi_S)$
asymmetry is of the size of a few percent and can be measured
through the experiment. The $\sin(3\phi-\phi_S)$ asymmetry is
smaller than the $\sin(\phi+\phi_S)$ asymmetry. After a cut on
$q_T$, we succeed in enhancing the asymmetry.

\end{abstract}

\end{frontmatter}
\section{introduction}
The single transverse spin asymmetries (SSAs) appearing in various
high energy scattering processes~\cite{bdr,Barone:2010ef} are among
the most interesting issues of QCD spin physics. Substantial SSAs in
semi-inclusive deeply inelastic scattering
(SIDIS)~\cite{smc,Airapetian:2004tw,compass,hermes05,compass06,2009ti,Alekseev:2010rw},
with one colliding nucleon transversely polarized, have been
measured by several experiments. These asymmetries, together with
the large asymmetries measured in $p p^\uparrow \rightarrow\pi X$
process~\cite{Adams:1991rw,Adams:1991cs,Adams:2003fx}, cannot be
explained by the leading-twist collinear picture of
QCD~\cite{Kane:1978nd}. It is found that the time-reversal-odd
($T$-odd) distribution functions or fragmentation functions play
essential roles for these asymmetries. Of particular interests, are
the leading-twist $T$-odd transverse momentum dependent (TMD)
distribution functions, such as the Sivers function~
\cite{sivers,anselmino95} and the Boer-Mulders function~\cite{bm}.
They arise from the correlation between the nucleon/quark transverse
spin and the quark transverse momentum, and can provide necessary
interference of amplitudes with different helicities and phases for
SSAs in the leading twist. Hence the study on these TMD
distributions and associated asymmetries can provide new insights
into QCD dynamics and nucleon
structure~\cite{bhs02,collins02,belitsky,Boer:2003cm}.

In this Letter we will explore the leading-twist SSAs in $\pi
p^\uparrow$ Drell-Yan process where the proton is transversely
polarized. In the TMD factorization picture, there are two
mechanisms for the SSAs in $\pi p^\uparrow$ Drell-Yan process at the
leading twist. One is the Sivers effect from the transversely
polarized proton, characterized by the $\sin(\phi-\phi_S)$ angular
dependence, where $\phi$ and $\phi_S$ are the azimuthal angles of
the lepton pair and the proton transverse spin, respectively. The
other one is the Boer-Mulders effect from the $\pi$ meson. The
studies on the former one were carried out in
Refs.~\cite{Efremov:2004tp,Anselmino2009}, which revealed the
possibility to test the sign reversal of $T$-odd distributions
between SIDIS and Drell-Yan process, a crucial prediction of QCD
dynamics. Here we will explore the SSAs in $\pi p^\uparrow$
Drell-Yan process from the Boer-Mulders effect of the $\pi$ meson,
since they have not been studied in detail phenomenologically. As
Boer-Mulders function is chiral-odd, it can only convolute with
another chiral-odd object constrained by helicity conservation in
hard partonic scattering process. For a transversely polarized
proton the leading-twist chiral-odd structure is manifested by the
transversity distribution and the pretzelosity distribution, of
which the combinations with the Boer-Mulders function yield the
$\sin (\phi+\phi_S)$ and $\sin(3\phi-\phi_S)$ azimuthal asymmetries,
respectively. The transversity distribution, usually denoted as
$h_1(x)$, can be interpreted as the difference between the densities
of quarks with transverse (Pauli-Lubanski) polarization parallel or
anti-parallel to the transverse polarization of the nucleon. Due to
its chiral-odd nature, transversity cannot be measured in inclusive
DIS process. Other than the double transversely polarized Drell-Yan
process, it was realized that transversity can also be accessed in
semi-inclusive DIS through the Collins effect~\cite{collins93}, and
in Drell-Yan process through the Boer-Mulders
effect~\cite{Boer1999}. The pretzelosity distribution, denoted as
$h_{1T}^\perp(x,\boldsymbol{p}^2_T)$, provides supplementary
chiral-odd structure of transversely polarized nucleon, especially
when the parton transverse momentum is probed. Studies on
pretzelosity can be found in Refs.~\cite{pretzelosity,Shejun2009}.
Further interest in pretzelosity relies on the observation that it
provides the information of the quark orbital angular momentum
inside the nucleon~\cite{Shejun2009} in a model dependent manner.
Besides the SIDIS process, pretzelosity can also be accessed in
single polarized Drell-Yan process.

In the present work, we study the SSAs in $\pi p^\uparrow$ Drell-Yan
process contributed by the coupling of the Boer-Mulders function of
the pion with the transversity and pretzelosity of the nucleon,
respectively. The hadron program by the COMPASS collaboration will
start at CERN, in which a $\pi^-$ beam colliding with proton target
is going to be available. In this work, we estimate the $\sin
(\phi+\phi_S)$ and $\sin(3\phi-\phi_S)$ azimuthal asymmetries at
COMPASS, not only for the $\pi^- p^\uparrow$ process, but also for
the $\pi^+ p^\uparrow$ process. We show that $\pi p^\uparrow$
Drell-Yan process could be applied to probe the chiral-odd structure
of the transversely polarized nucleon in the leading twist.

\section{Single spin asymmetry in Drell-Yan process}
For a general Drell-Yan process with one of the beam transversely
polarized, i.e., $h_1 h_2^\uparrow\rightarrow \ell^+ \ell^- X$, the
single spin asymmetry is simply defined as
\begin{eqnarray}
A_{UT}=\frac{d\sigma^{h_1 h_2^\uparrow\rightarrow \ell^+ \ell^- X} -
d\sigma^{h_1 h_2^\downarrow\rightarrow \ell^+ \ell^-
X}}{d\sigma^{h_1 h_2^\uparrow\rightarrow \ell^+ \ell^- X} +
d\sigma^{h_1 h_2^\downarrow\rightarrow \ell^+ \ell^- X}} \equiv
\frac{d\sigma^\uparrow - d\sigma^\downarrow}{d\sigma^\uparrow +
d\sigma^\downarrow}.
\end{eqnarray}
Here we treat the process in the parton model and only consider the
leading order approximation via a single photon transfer, i.e.,
$q\bar{q}\rightarrow\gamma^*\rightarrow\ell^+\ell^-$. We denote the
momenta of the hadrons, the annihilating partons, and the produced
lepton pairs as $P_i$, $p_i$ and $k_i~(i=1,2)$, respectively. Then
the momentum transfer gives the invariant mass of the lepton pair
\begin{eqnarray}
q^2=(p_1+p_2)^2=(k_1+k_2)^2=M^2.
\end{eqnarray}
Now we work in the center of mass frame of two hadrons, and
parameterize the four-momentum of the photon as
$q=(q_0,\bm{q}_T,q_L)$. At extremely high energies, if we assume
that the longitudinal component is dominant and neglect all the mass
effects and the transverse momentum, we can define the following
variables,
\begin{eqnarray}
&&x_1=\frac{q^2}{2P_1\cdot
q}\approx\frac{q_0+q_L}{\sqrt{s}},~~~~x_2=\frac{q^2}{2P_2\cdot
q}\approx\frac{q_0-q_L}{\sqrt{s}},\nonumber\\
&&\tau=\frac{M^2}{s},~~~~x_F=x_1-x_2\approx\frac{2q_L}{\sqrt{s}}.
\end{eqnarray}
Then we can build up the relation
\begin{eqnarray}
&&x_1=\frac{1}{2}\big(x_F+\sqrt{x_F^2+4\tau}\big),\nonumber\\
&&x_2=\frac{1}{2}\big(-x_F+\sqrt{x_F^2+4\tau}\big).\label{x}
\end{eqnarray}

The direction of the detected lepton pair can be described by the
solid angle ($\theta,\phi$), which is frame dependent. In our Letter,
we will select the Collins-Soper frame~\cite{CS_frame}. The
convention for the definition of the azimuthal angles of the lepton
pair and proton transverse spin in our Letter is the same as that used in
Refs.~\cite{Boer1999,Bacchetta2010}, but different with that in
Ref.~\cite{Arnold2009}, although we can easily demonstrate that they
are equivalent by a simple transformation.

Our aim is to explore the transversity and pretzelosity through SSA
in $\pi p^\uparrow$ Drell-Yan process, so we write down the cross-section~\cite{Boer1999,Bacchetta2010,Arnold2009} only with the terms
we are interested in:
\begin{eqnarray}
\label{cross_section}
&&\frac{d\sigma}{d\Omega dx_1 dx_2
d^2\bm{q}_T}=\frac{\alpha^2}{3q^2}\{A(y)\mathcal{F}[\bar{f}_1f_1]\nonumber\\
&&-|\bm{S}_{2T}|[B(y)\sin(\phi+\phi_{S_2})\times\mathcal{F}
[\frac{\hat{\bm{h}}\cdot\bm{p}_{1T}}{M_1}\bar{h}_1^\perp h_1]
+B(y)\sin(3\phi-\phi_{S_2})\nonumber\\
&&
\times\mathcal{F}[\frac{4\hat{\bm{h}}\cdot\bm{p}_{1T}(\hat{\bm{h}}\cdot\bm{p}_{2T})^2
-2\hat{\bm{h}}\cdot\bm{p}_{2T}\bm{p}_{1T}\cdot\bm{p}_{2T}
-\hat{\bm{h}}\cdot\bm{p}_{1T}\bm{p}_{2T}^2}{2M_1M_2^2}\bar{h}_1^\perp
h_{1T}^\perp]]\}\nonumber\\
&&+...,
\end{eqnarray}
where the notation $\mathcal{F}$ is defined as
\begin{eqnarray}
\mathcal{F}[\omega \bar{f} g]&\equiv&\sum_{a,\bar{a}}\int
d\bm{p}_{1T}d\bm{p}_{2T}
\delta^2(\bm{p}_{1T}+\bm{p}_{2T}-\bm{q}_{T})\omega(\bm{p}_{1T},\bm{p}_{2T})\nonumber\\
&\times& \bar{f}^{\bar{a}}(x_1,\bm{p}_{1T})g^{a}(x_2,\bm{p}_{2T}),
\end{eqnarray}
and
\begin{eqnarray}
A(y) =  \frac{1}{2} - y + y^2  \stackrel{\mbox{cm}}{=}\frac{1}{4}( 1
+ \cos^2 \theta),~~~~~B(y) = y (1-y) \,
\stackrel{\mbox{cm}}{=}\frac{1}{4} \sin^2 \theta  .
\end{eqnarray}
The parton distribution functions (PDFs) which appear in the
expression are transverse momentum dependent (TMD) parton
distributions, whose factorization in SIDIS and Drell-Yan processes
has been proved in
Refs.~\cite{Collins1985,Ji2004,Collins:2004nx,Ji2005}, and holds in
the regime $q_T^2\ll M^2,~p_i\simeq q_T$. Similar to the SIDIS
process, the integration over $\bm{q}_T$ directly leads to zero
result for the last two terms in Eq.~(\ref{cross_section}), so we
have to define the weighted asymmetry
\begin{eqnarray}
A^{W(\phi,\phi_S)}_{UT}=\frac{2\int_0^{2\pi}d\phi
W(\phi,\phi_S)[d\sigma^\uparrow -
d\sigma^\downarrow]}{\int_0^{2\pi}d\phi [d\sigma^\uparrow +
d\sigma^\downarrow]}.
\end{eqnarray}
With proper weight functions $W(\phi,\phi_S)$ which depend on the
azimuthal angles $\phi$ and $\phi_S$, we can distinguish the
$\sin(\phi+\phi_S)$ and the $\sin(3\phi-\phi_S)$ asymmetries from
the differential cross-section. In this study, we will plot the
$\sin(\phi+\phi_S)$ and the $\sin(3\phi-\phi_S)$ weighted
asymmetries, for they can give us the information on the
transversity and pretzelosity we are interested in. Also we remind that a Monte Carlo
simulation for the $\sin(\phi+\phi_S)$ asymmetry in $\pi p^\uparrow$
Drell-Yan process has been given in~\cite{Bianconi:2006hc}.

\section{Numerical calculation}

We will consider the $\pi^-(\pi^+) p^\uparrow\rightarrow\mu^+\mu^-X$
process, where a valence $\bar{u}(\bar{d})$ quark from
$\pi^-(\pi^+)$ and a $u(d)$ quark from $p$ annihilate. Here we
have ignored the contribution from sea quarks since we assume that
the polarized effect from sea quarks is small. For the proton PDFs,
we will use the results obtained in a light-cone quark spectator
diquark model~\cite{Ma,Shejun2009} with the relativistic
Melosh-Wigner effect~\cite{MaOld} of quark transversal motions taken
into account. The TMDs we deduce from this model are applicable in
the hadronic scale. To compare with the experimental observables
which are usually measured at rather high energies, it is essential
to evolve the parton distributions to the scale from an initial
scale. However, here we calculate the azimuthal asymmetries which
are the ratios of different parton distributions, so the effects of
evolution are assumed to be small. In practice, we use this model to
obtain the helicity and transversity distributions, which are
reasonable to describe data related to helicity distributions in a
number of processes~\cite{Chen2005} and transversity distributions
related to the Collins asymmetry at HERMES~\cite{Huang2007}.  This
model is also successful in the prediction of the dihadron
production asymmetry at COMPASS~\cite{She2008,compass_dihadron}. So
it is worth trying to apply this model to the Drell-Yan kinematics
at COMPASS. Besides this, we need the Boer-Mulders function for a
pion~\cite{Lu,Gamberg:2009uk}, of which the knowledge is limited,
and we will use the parametrization in Ref.~\cite{Lu}, which was
obtained in a quark spectator antiquark model. The pion parton
distributions we adopt were demonstrated~\cite{Lu} to give a good
description on the $\cos 2\phi$ asymmetries measured in the
unpolarized $\pi N$ Drell-Yan process~\cite{NA10}, where a large and
increasing asymmetry was observed in the $q_T$ region below 3 GeV,
thus our model has been checked to be reasonable in this region.
Another important feature we should remember is that this $T$-odd
function has a different sign in the Drell-Yan process with that in
the SIDIS process~\cite{collins02,Boer:2003cm,Bomhof2008},
\begin{eqnarray}
h_1^\perp|_{\mathrm{DY}}=-h_1^\perp|_{\mathrm{SIDIS}}.
\end{eqnarray}
In Ref.~\cite{Lu}, the Boer-Mulders function is calculated for the
SIDIS process, so we will make a sign change for our parametrization
in our Letter. However, we should be careful that due to the
chiral-odd nature of Boer-Mulders function, it always couples with
another chiral-odd function for being probed. This makes it very
difficult to obtain the information of this function, especially its
sign. In the unpolarized Drell-Yan process, the Boer-Mulders
function couples with itself, therefore it is impossible to
determine its sign. In the SIDIS process, the Boer-Mulders function
is combined with the Collins function, the
extraction~\cite{Barone:2008tn} of which also relies on the
azimuthal asymmetry of hadron production in $e^+ e^-$ annihilation
process. Also we will stress that unlike many other calculations, we
do not make the ansatz that the transverse momentum dependence of
the TMDs has a pure Gaussian form, but just deduce it from the
model. That is, we evaluate the integration over the parton
transverse momenta numerically.  The experiment we consider is for
COMPASS, where the kinematics we will use are~\cite{Takekawa2010}
\begin{eqnarray}
&&\sqrt{s}=18.9~\mathrm{GeV},~0.1<x_1<1,~0.05<x_2<0.5,\nonumber\\
&&4\leqslant M\leqslant8.5~\mathrm{GeV},~0\leqslant q_T \leqslant
4~\mathrm{GeV}~(\mathrm{if}~q_T~\mathrm{is~integrated}).\nonumber
\end{eqnarray}

\begin{figure}
\includegraphics[scale=0.7]{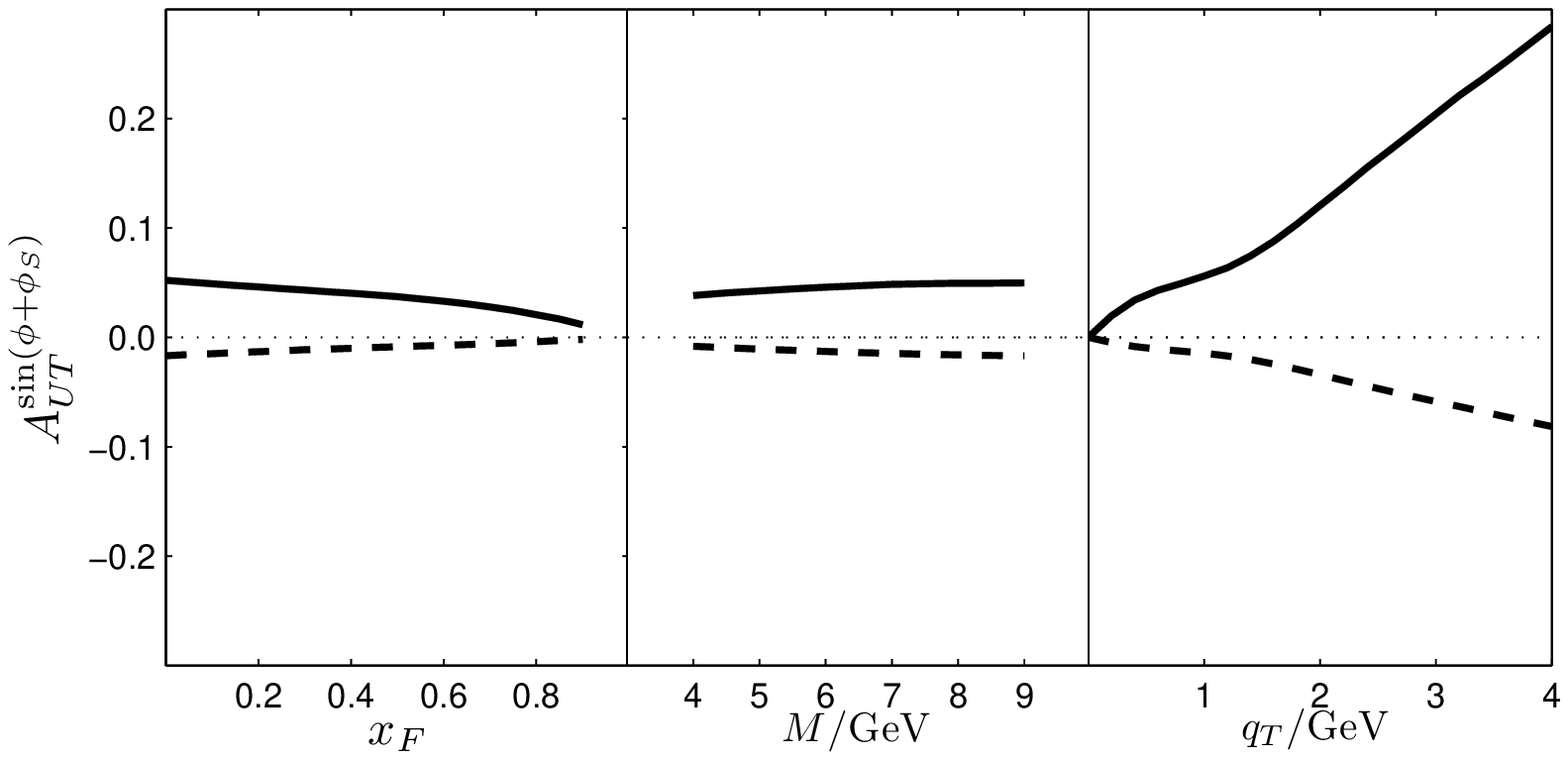}
\caption{The $\sin(\phi+\phi_S)$ asymmetries for $\pi^{\pm}
p^\uparrow\rightarrow\mu^+\mu^-X$ process at COMPASS. Solid and
dashed curves are the results for $\pi^-$ and $\pi^+$ productions,
respectively.} \label{tran}
\includegraphics[scale=0.7]{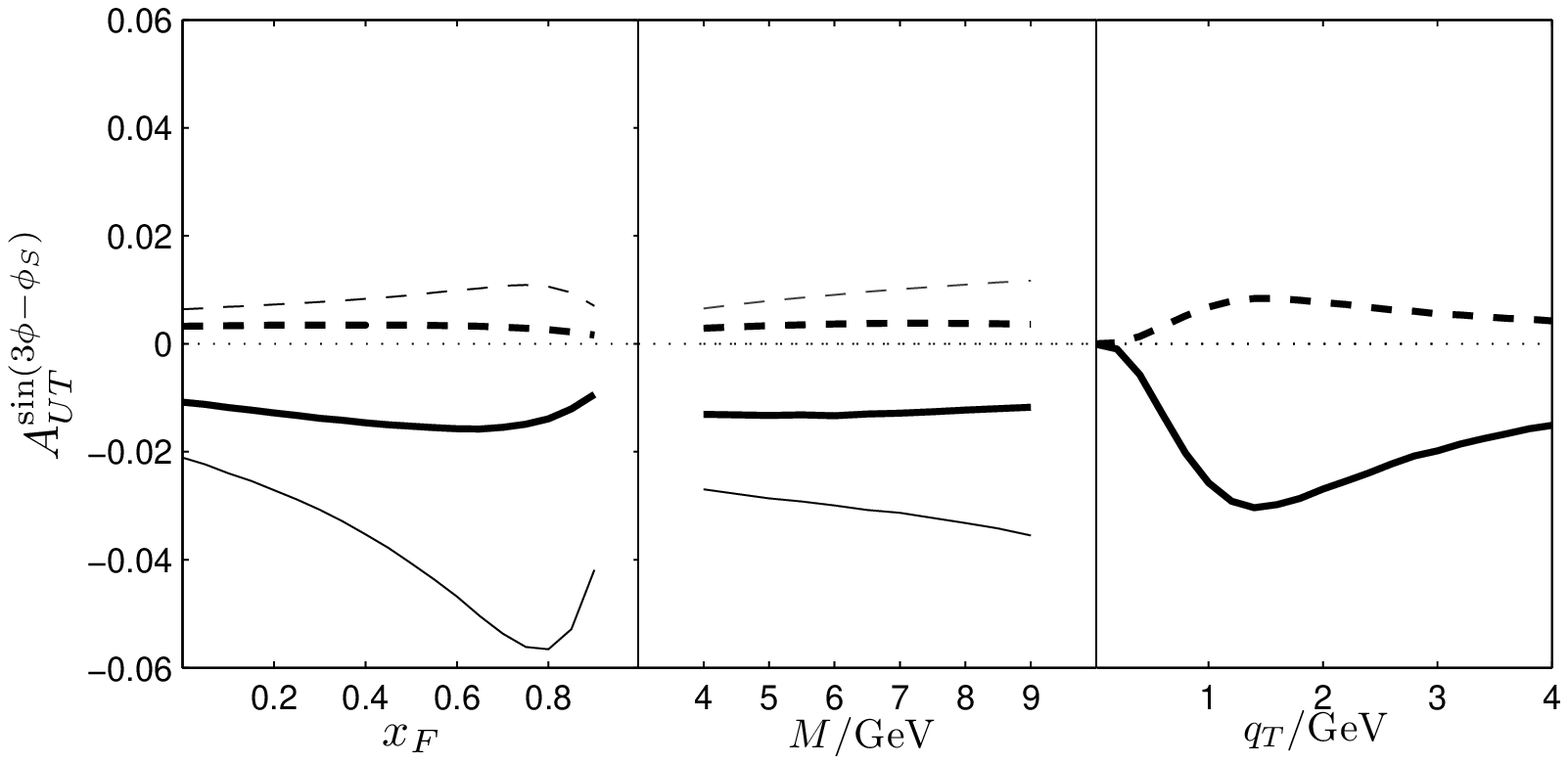}
\caption{Similar to Fig.~\ref{tran}, but for the
$\sin(3\phi-\phi_S)$ asymmetries. The thin curves are calculated
with a cut $1.0\leqslant q_T \leqslant 2.0$~GeV.} \label{pret}
\end{figure}

We will investigate the $x_F, M$ and $q_T$ dependence of the
asymmetries. The integration range can be determined as follows.
\begin{itemize}
\item For the $x_F/M$ dependence, given a fixed $x_F/M$, the range for $M/x_F$ is
determined by Eq.~(\ref{x}) so that
$x_{1,2}^{\mathrm{min}}<x_{1,2}(x_F,M)<x_{1,2}^{\mathrm{max}}$.
\item For the $q_T$ dependence, the range for $M$ is $4\leqslant
M\leqslant8.5~\mathrm{GeV}$, and the range for $x_F$ is determined
by Eq.~(\ref{x}) so that
$x_{1,2}^{\mathrm{min}}<x_{1,2}(x_F,M)<x_{1,2}^{\mathrm{max}}$.
\end{itemize}
In Figs.~\ref{tran} and Fig.~\ref{pret} (thick curves), we plot the
$\sin(\phi+\phi_S)$ asymmetry and the $\sin(3\phi-\phi_S)$ asymmetry
in the $\pi p^\uparrow$ Drell-Yan at COMPASS, respectively.
We can clearly see from the two figures that the asymmetries for the
$\pi^- p^\uparrow$ process are much larger than those for the $\pi^+
p^\uparrow$ process, because that the former process is dominated by
$u$ quark while the latter is dominated by $d$ quark. COMPASS will
conduct a $\pi^- p^\uparrow$ plan in the near future, however, we
will also give the prediction on the $\pi^+ p^\uparrow$ process as a
supplement, and expect future experiments could direct this
measurement to give us more information on the $d$ quark
distributions.

From Fig.~\ref{tran}, we find that similar to that in the SIDIS
process, the $\sin(\phi+\phi_S)$ asymmetry is also significant in
the Drell-Yan process, and the magnitude of the asymmetry reaches up
to several percent. We can make a comparison with the results from
Ref.~\cite{Boer2008}, where a $q_T$ dependence of the
$\sin(\phi+\phi_S)$ asymmetry was investigated. Below 2~GeV, our two
results seem to be consistent, but above 2~GeV, our result rises
quickly and give a much larger asymmetry than that obtained in
Ref.~\cite{Boer2008}. This needs further studies and a check by
experiments. As to the $\sin(3\phi-\phi_S)$ asymmetry, however, we
are not fortunate that it is not so large, but just around $1-2\%$,
also similar to that shown in the SIDIS process~\cite{Shejun2009}.
In order to enhance the asymmetry, it has been suggested in
Ref.~\cite{Shejun2009} to make a cut on $P_{h\perp}$ by selecting
the large $P_{h\perp}$ events. But there we faced a dilemma that
$P_{h\perp}$ cannot be too large to spoil the TMD factorization,
which only holds at the regime $P_{h\perp}\ll Q^2$. Here we will
adopt the same approach to try to enhance the asymmetry in Drell-Yan
process, i.e., we will make a cut on $q_T$. From the third subplot
in Fig.~\ref{pret}, we find that the asymmetry is larger in the
medium $q_T$ region. So we choose the cut $1.0\leqslant q_T
\leqslant 2.0$~GeV, and this kinematics region on $q_T$ also
satisfies the condition $q_T^2\ll M^2$, thus the TMD factorization
is still valid. Without changing other kinematics and just
integrating $q_T$ from 1.0~GeV to 2.0~GeV, we recalculate the
$\sin(3\phi-\phi_S)$ asymmetry, and the result is shown in
Fig.~\ref{pret} (thin curves). As we expect, the magnitude of the
asymmetries in indeed is enhanced by about two times after we make a
cut on $q_T$. Although we may have a loss on the data, we hope that
it would be helpful to measure this asymmetry from experiments. Here
we emphasize that the weighting functions we choose depend only on
$\phi$ and $\phi_S$, but not on the transverse momenta of the
dilepton. Meanwhile, we perform the integration over the parton
transverse momenta numerically, therefore we do not need to
introduce the transverse moments of the TMDs in our calculation.

In the SIDIS process, if we want to extract transversity or
pretzelosity, we should know the information about the Collins
fragmentation function. In the Drell-Yan process, transversity or
pretzelosity also needs a chiral-odd partner, e.g., the Boer-Mulders
function, of which the knowledge is limited, especially for a pion
beam. However, the Boer-Mulders function of the pion can be accessed
through unpolarized pion nucleon Drell-Yan process at
COMPASS~\cite{Lu:2006ew}. Furthermore, all the TMDs we used in our
Letter are from the same model, therefore they are consistent with
each other. Thus the relevant experiment will give constraints on
the transversity/pretzelosity and Boer-Mulders functions, though a
complete knowledge on the TMDs must rely on more experiments.

\section{Conclusion}
We have presented the $\sin(\phi+\phi_S)$ and $\sin(3\phi-\phi_S)$
single spin asymmetries for the $\pi^\pm
p^\uparrow\rightarrow\mu^+\mu^-X$ process at COMPASS. For the $\pi^-
p^\uparrow$ process, the $\sin(\phi+\phi_S)$ asymmetry is several
percent and can be measured through the experiment. However, the
$\sin(3\phi-\phi_S)$ asymmetry is small, which is similar to the
case in the SIDIS process, thus there is some difficulty in
measuring it. We adopt a cut on $q_T$ as used before to solve the
similar difficulty in SIDIS process, and our attempt succeeds in
enhancing the asymmetry. For the $\pi^+ p^\uparrow$ process, we get
an expected smaller result due to the different quark dominance. Our
purpose is to study transversity and pretzelosity of the proton
through SSAs in Drell-Yan process, for this we apply the
Boer-Mulders effect of a pion beam, which will be available at
COMPASS. Therefore our predictions on the $\sin(\phi+\phi_S)$ and
$\sin(3\phi-\phi_S)$ asymmetries in Drell-Yan process rely on the
knowledge of the Boer-Mulders function. Nevertheless, our model
prediction can give constraints on the relevant physical quantities,
and we expect more experiment data to provide us more knowledge on
the spin structure of the nucleon, especially, the chiral-odd
structure of the transversely polarized nucleon.

\section*{Acknowledgement}
This work is partially supported by National Natural Science
Foundation of China (Nos.~10905059, 11005018, 11021092, 10975003,
11035003) and by FONDECYT (Chile) under project No.~11090085.

\end{document}